\documentclass[twocolumn,showkeys,preprintnumbers,amsmath,amssymb]{revtex4}
\usepackage{graphicx}
\usepackage{dcolumn}
\usepackage{bm}

\begin{document}


\title{Goodness of Generalized Seniority in Semi-magic Nuclei}
\author{ASHOK KUMAR Jain \footnotetext{ \textbf{Foundation item:} Ministry of Human Resource and Development, Govt. of India.\\ 
\textbf{Biography:} ASHOK KUMAR Jain, male (Indian), Varanasi, India, Professor, Working on nuclear structure physics}}
\email{ajainfph@iitr.ac.in}

\author{BHOOMIKA Maheshwari}
\affiliation{Department of Physics, Indian Institute of Technology, Roorkee 247667, India.}

\begin{abstract}
Symmetry plays an important role in understanding the nuclear structure properties from the 
rotation of a nucleus to the spin, parity and isospin of nuclear states. This simplifies the complexity of the nuclear problems in one way or the other. Seniority is also a well known quantum number which arises due to the symmetry in the pairing interaction of nuclei. We present empirical as well as theoretical evidences based on decay rates which support the goodness of seniority at higher spins as well as in n-rich or, n-deficient nuclei. We find that the generalized seniority governs the identical trends of high-spin isomers in different semi-magic chains, where different set of nucleon orbitals from different valence spaces are involved.     
\end{abstract}


\maketitle                              
\section{Introduction}

Understanding the complex nuclear structure of the atomic nucleus is an outstanding problem of nuclear physics. The pairing of nucleons is well known for several decades, which enables one in describing the  ground states of even-even nuclei, and the extra stability of nuclei with an even number of nucleons than the nuclei with an odd number, etc. In atomic physics, pairing was understood in terms of seniority, first introduced by Racah in 1940s~\cite{Racah1943}. In 1950s, Racah and Talmi~\cite{Racah1952}, and Flowers~\cite{Flowers1952} independently introduced the seniority scheme in nuclear physics. In simple terms, seniority may be defined as the number of unpaired nucleons in a given state, generally denoted as $v$. It is now known that $v$ remains a good quantum number for states emanating from a pure-j configuration with j$\le$7/2. However, its validity for higher-j values has also been suggested. 

Arima and Ichimura~\cite{Arima1966}, and Talmi~\cite{Talmi1971} further extended the seniority picture in single-j to the generalized seniority in multi-j. Generalized seniority takes care of the presence of multi-j orbitals in a given state. Semi-magic nuclei provide a fertile ground to study the various properties on the basis of the seniority scheme. One of the most interesting results from this scheme is the formation of seniority isomers. Seniority isomers are one of the well known categories of the nuclear isomers, i.e. the longer lived excited states, where the hindrance to their decays have been explained in terms of seniority selection rules~\cite{Jain2015}. It has been a general belief that the seniority isomers arise only in $E2$ decays between the same seniority states due to the vanishing decay probabilities at the mid-shell in seniority scheme. We have recently used the simple quasi-spin scheme to obtain the generalized seniority results in a mixed configuration coming from several degenerate orbitals, and applied it to the high-spin isomers in Sn isotopes~\cite{Maheshwari2016}. Hence, we have found for the first time odd-tensor $E1$ decaying ${13}^-$ isomers in Sn isotopes, a new category of isomers. We have then used the same scheme to understand the first excited $2^+$ states in Sn isotopes and explained the asymmetric twin $B(E2)$ parabolas~\cite{Maheshwari20161}. 
    
In this paper, we have applied the ``generalized seniority formalism'' to the high-spin nuclear isomers in the semi-magic nuclei, particularly Z=50 isotopes, N=82 isotones and Z=82 isotopes. We find that the generalized seniority remains a reasonably good quantum number for a set of states, particularly the high-spin isomers in these semi-magic nuclei. This further governs the identical behavior of these isomeric states in different semi-magic chains, having different set of active orbitals. We start by presenting a few empirical evidences and understanding the reasons behind, in section 2, which makes a good ground for the seniority calculations. Thereafter, we follow these empirical findings with our generalized seniority calculations and results for both even- and odd-A semi-magic nuclei. The overall conclusions of the present work have been presented in the last section.

\section{Experimental evidences}

The level schemes of the $^{119-130}$Sn isotopes have been studied by using the reactions induced by light ions, deep inelastic reactions, or fission fragment studies by several researchers~\cite{daly80, fogelberg81, daly86, lunardi87, broda92, mayer94, daly95, pinston00, zhang00, lozeva08}. Many isomer systematics have been identified for $N>64$ Sn-isotopes, and the isomeric states ${10}^+$ and ${27/2}^-$ have been characterized as seniority $\it{v}$=2 and $\it{v}$=3 states in these studies, where $\it{v}$ denotes the seniority quantum number. Pietri et al.~\cite{pietri11} recently identified and confirmed the high-spin and high-seniority $\it{v}$=4, ${15}^-$ isomeric state in $^{128}$Sn. More recently, Astier et al.~\cite{astier132, astier125} reported detailed high-spin level schemes in the $^{119-126}$Sn isotopes by using the binary fission fragmentation induced by heavy ions. Iskra et al.~\cite{iskra14} have also focused on high-seniority states in neutron-rich, even-even Sn-isotopes. It may be noted that there exists some deformed collective states giving rise to a full or a part of rotational band in the even-even light mass Sn isotopes with $A=110-118$, interpreted as $2p-2h$ proton configuration~\cite{bron79, poelgeest80, harada88, savelius98, gableske01, wolinska05, wang10}. But the ${10}^+$ yrast isomeric states discussed in the present paper are not part of any rotational structure~\cite{fotiades11}. More recently, the studies on Sn isotopes have been pushed much beyond the $N=82$ shell closure and isomers in the $N=86-88$ Sn-isotopes have been populated by Simpson et al.~\cite{simpson14} which shed a new light on the effective interaction in n-rich nuclei~\cite{maheshwari15}. 

The ${10}^+$ and ${27/2}^-$ isomers have also been identified, in the $N=82$ isotonic chain from $Z=66$, Dy to $Z=72$, Hf, as seniority $\it{v}$=2 and $\it{v}$=3 isomers coming from the $h_{11/2}$ proton orbital~\cite{mcneill89}. Recently, the high-spin structure of five $N=82$ isotones with $Z=54-58$ has also been reported by Astier et al.~\cite{astier126}, where the ${10}^+$ isomers have been described as broken pairs of protons from the $g_{7/2}$ and $d_{5/2}$ orbitals in the even-mass isotones.

We plot the excitation energies of the ${10}^+$ isomers relative to $0^+$ states and the ${27/2}^-$ isomers relative to ${11/2}^-$ states for the $Z=50$ isotopes and the $N=82$ isotones in the top and bottom panels of Fig. 1, respectively. It may be noted that the same valence orbitals are involved in both the $Z=50$ and $N=82$ chains. While the neutrons occupy these orbitals in the $Z=50$ isomers, the protons take over the role in the $N=82$ isomers. We find that all the main features observed in the $Z=50$ isotopic chain are also present in the $N=82$ isotonic chain and both appear to be nearly identical to each other. We can see that the energy gap is almost constant and particle number independent which is a well known signature of nearly good seniority ~\cite{talmi93, lawson80, talmi03}. The ${10}^+$ and the ${27/2}^-$ isomers, belonging to the even-even and even-odd nuclei respectively, are seen to follow each other very closely throughout the chains, if one puts the $0^+$ and ${11/2}^-$ states on equal footing. This suggests that the nuclear configurations and structure for the ${10}^+$ and the ${27/2}^-$ isomers should be very similar without any odd-even effect which also suggests the aligned nature of the involved nucleons.

\begin{center}
\includegraphics[width=8cm,height=6cm]{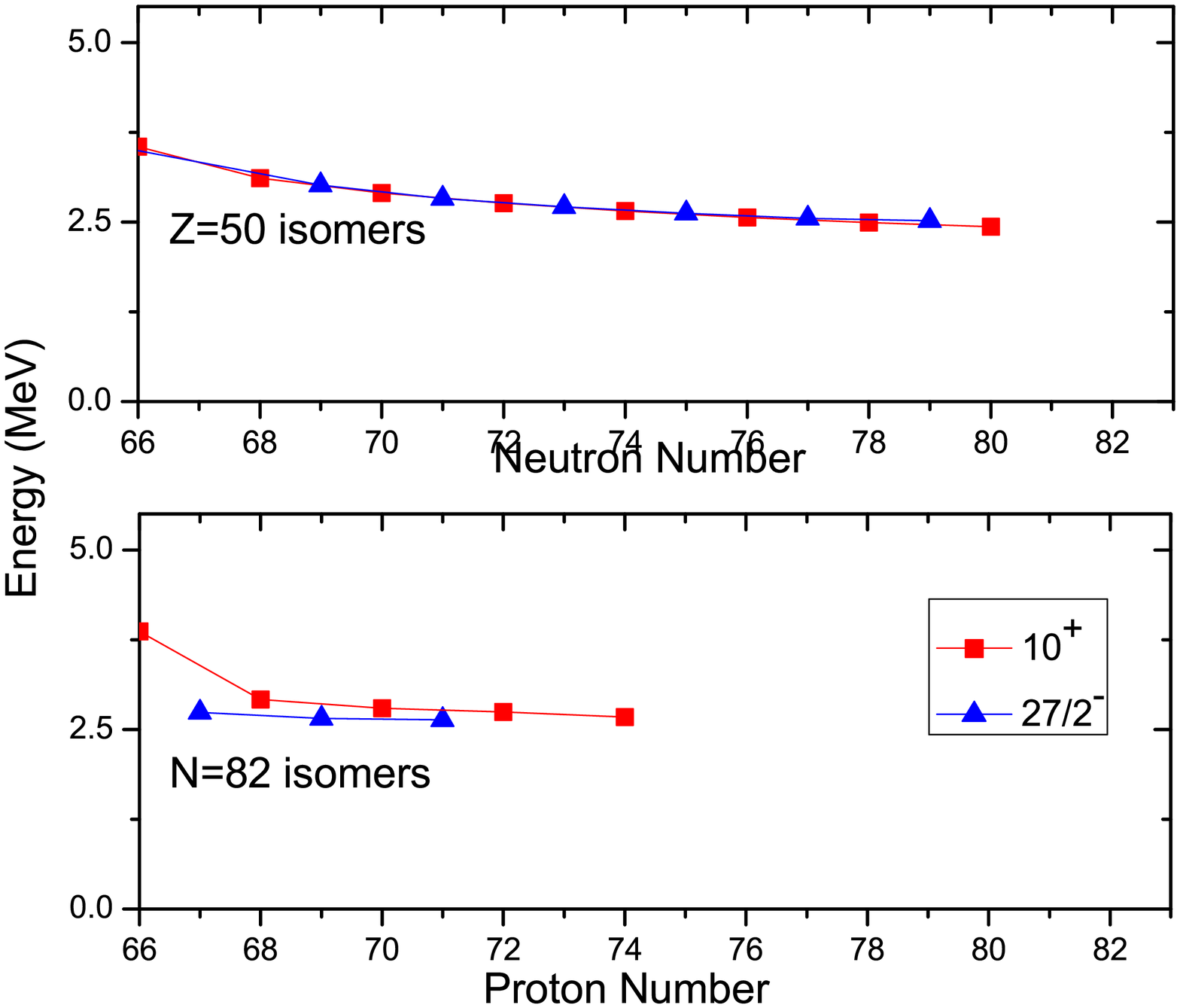}\\
Fig.\ 1\begin{minipage}[t]{69mm} \quad (color online)\,Variation of the experimental energy values of the ${10}^+$ and ${27/2}^-$ isomers in $Z=50$ isotopes and $N=82$ isotones. \\
\end{minipage}
\label{fig:energyz50n82}
\end{center}
We have plotted the measured half-lives (in $\mu$s) of these isomers with increasing nucleon numbers in the top and bottom panels of Fig. 2, for the $Z=50$ and $N=82$ chains, respectively. The half-lives of the ${10}^+$ and ${27/2}^-$ isomers exhibit a rise near the middle of the active valence space (from neutron/proton numbers $66$ to $82$), attain a maximum value, and fall with increasing nucleon number. The ${10}^+$ and ${27/2}^-$ isomeric states, for the $Z=50$ isotopes, exhibit a maximum at the neutron numbers $72$ and $73$ respectively, where the $h_{11/2}$ neutron orbital becomes half-filled~\cite{mayer94, zhang00, lozeva08}. On the other hand, for the $N=82$ isomers, the peaks are observed at $Z=70$ and $71$ for the ${10}^+$ and ${27/2}^-$ isomeric states respectively, where the $h_{11/2}$ proton orbital becomes half-filled~\cite{mcneill89}. This happens because the electric quadrupole (E2) transition probabilities between a state $J_i$ and another state $J_f$ with same seniorities vanish at the middle~\cite{talmi93,lawson80,Maheshwari2016}. We can, therefore, foresee that the isomeric half-lives at the middle of the active valence space in the $Z=50$ and $N=82$ chains are most affected by the seniority selection rules. We also notice that the half-lives of odd-A Sn isotopes, i.e. the ${27/2}^-$ isomers are lower than the neighboring even-A Sn isotopes, i.e. the ${10}^+$ isomers as expected from pairing consideration.
\begin{center}
\includegraphics[width=8cm,height=6cm]{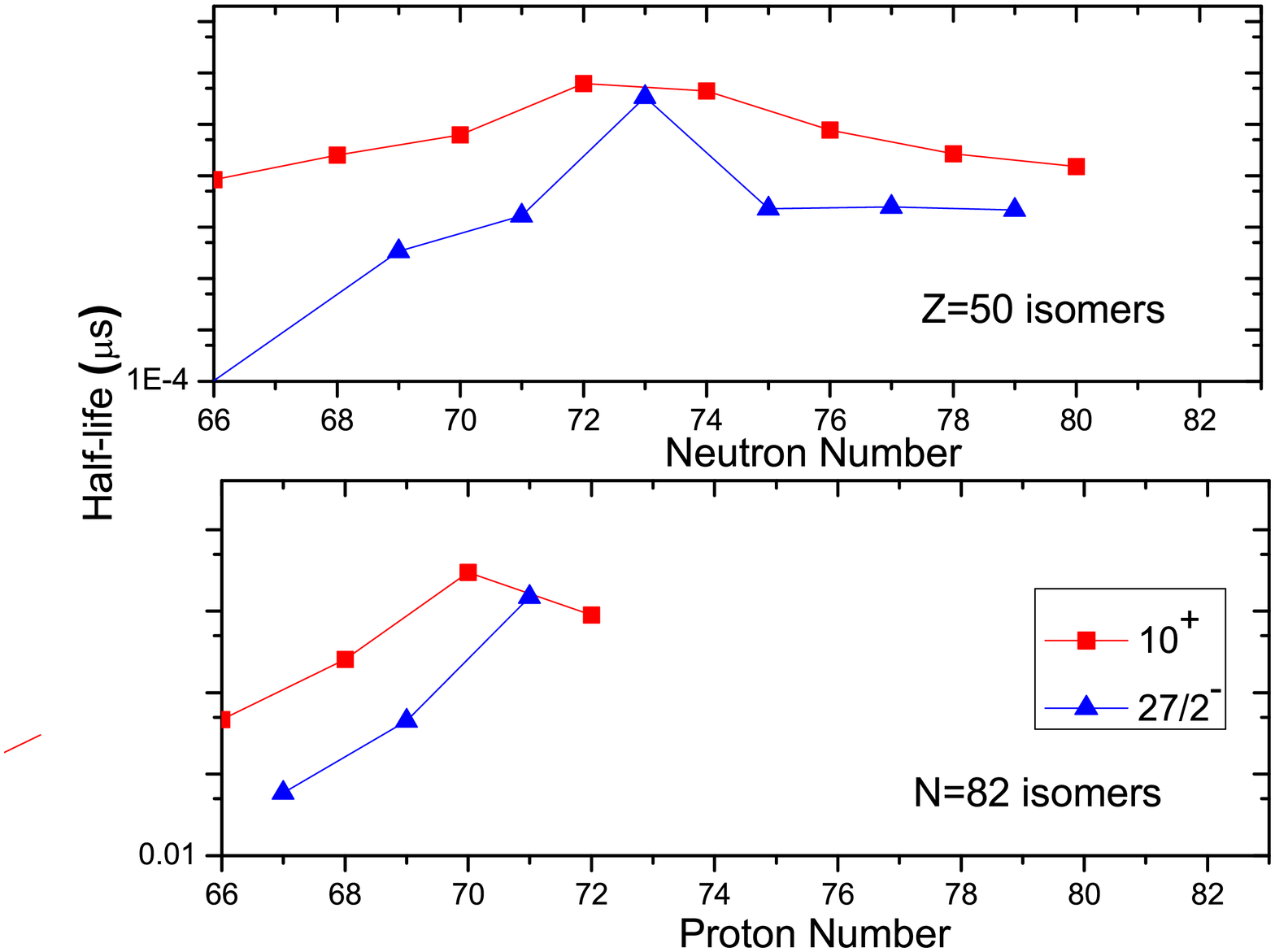}\\
Fig.\ 2\begin{minipage}[t]{69mm} \quad (color online)\,Variation of the experimental half-life values of the ${10}^+$ and ${27/2}^-$ isomers in $Z=50$ isotopes and $N=82$ isotones. The vertical scale is logarithmic.\\
\end{minipage}
\label{fig:halflifez50n82}
\end{center}

We have also plotted the excitation energies of the $12^+$ isomers relative to $0^+$ states and the $33/2^+$ isomers relative to $13/2^+$ states in the top panel of Fig. 3 for the $Z=82$, Pb isotopes. All the experimental data in this paper have been adopted from our atlas~\cite{Jain2015}, the ENSDF (Evaluated Nuclear Structure Data File)~\cite{ENSDF}, and the XUNDL (Unevaluated Nuclear Data List)~\cite{XUNDL} data sets. We can again see that both the Pb-isomers, even-A and odd-A, closely follow each other, and do not show any odd-even effect, as in the cases of Sn isotopes and N=82 isotones. This again supports the empirical evidences of good seniority states with a particle number independent nature of the states. We, therefore, expect them to have similar origins in terms of their wave functions and nuclear configurations. We have also plotted their half-lives (in units of $\mu$s) in the bottom panel of Fig. 3, where one can observe that the half-lives show their increment towards mass number $A=190$ ($N=108$), which may be the middle of the active valence space for these Pb isomeric states.

\begin{table*}
Table\ 1 Comparison of the experimentally measured $\Delta E^{2^+}_{0^+}$ and $ \Delta E^{{15/2}^-}_{{11/2}^-}$ $\gamma-$transitions in even-even and odd-A Sn-isotopes for $N$ $\ge$ 66 and their ratio $R(15:2)$. Also, compared are the $\Delta E^{12^+}_{10^+}$ and $ \Delta E^{{31/2}^-}_{{27/2}^-}$ $\gamma-$transitions involving states which decay to the ${10}^+$, ${27/2}^-$ isomers and their ratio $R(31:12)$. All the energies are in MeV.\\
\begin{center}
\begin{ruledtabular}

\begin{tabular}{c c c c c c c c c}
\hline
Isotope & $ \Delta E^{2^+}_{0^+} $ & $\Delta E^{12^+}_{10^+}$ &Isotope & $\Delta E^{{15/2}^-}_{{11/2}^-}$ & $\Delta E^{{31/2}^-}_{{27/2}^-}$ &$R(15:2)$ & $R(31:12)$  \\
\hline
&&&&\\
$^{116}$Sn & ${1.294}$ &    & $^{117}$Sn & ${1.279}$ &   &${0.99}$&     \\
$^{118}$Sn & ${1.230}$ &1.237 &$^{119}$Sn & ${1.220}$ &1.179&${0.99}$&0.95\\
$^{120}$Sn & ${1.171}$ &1.190&$^{121}$Sn & ${1.151}$ &1.083&${0.98}$&0.91\\
$^{122}$Sn & ${1.141}$ &1.103&$^{123}$Sn & ${1.107}$ &1.043&${0.97}$&0.95\\
$^{124}$Sn & ${1.132}$ &1.047&$^{125}$Sn & ${1.088}$ &0.924&${0.96}$&0.88\\

\hline
\end{tabular}
\end{ruledtabular}
\label{tab:table1}
\end{center}
\end{table*}


\begin{table*}[htb]
Table\ 2 Same as Table 1, but for the $N=82$ isotones with Z $\geq$ 66. All the energies are in MeV.\\
\begin{center}
\begin{ruledtabular}

\begin{tabular}{c c c c c c c c}
\hline
Isotope & $ \Delta E^{2^+}_{0^+} $ & $\Delta E^{12^+}_{10^+}$ & Isotope & $\Delta E^{{15/2}^-}_{{11/2}^-}$ & $\Delta E^{{31/2}^-}_{{27/2}^-}$ & $R(15:2)$ & $R(31:12)$ \\
\hline
&&&&\\
$^{148}$Dy & ${1.677}$ &1.932&$^{149}$Ho & ${1.560}$ &   &${0.93}$&     \\
$^{150}$Er & ${1.578}$ &1.446&$^{151}$Tm & ${1.478}$ &1.332&${0.94}$&0.92\\
$^{152}$Yb & ${1.531}$ &     &$^{153}$Lu & ${1.432}$ &    &${0.94}$&\\
\hline
\end{tabular}
\end{ruledtabular}
\label{tab:table2}
\end{center}
\end{table*}
\begin{table*}[htb]
Table\ 3 Same as Table 1, but for the $Z=82$ isotopes with N $\leq$ 108. Comparison of the experimentally measured $\Delta E^{2^+}_{0^+}$ and $ \Delta E^{{17/2}^+}_{{13/2}^+}$ $\gamma-$transitions in even-even and odd-A Pb-isotopes and their ratio $R(17:2)$. All the energies are in MeV.\\
\begin{center}
\begin{ruledtabular}

\begin{tabular}{c c c c c c c c}
\hline
Isotope & $ \Delta E^{2^+}_{0^+} $ & $\Delta E^{14^+}_{12^+}$ & Isotope & $\Delta E^{{17/2}^+}_{{13/2}^+}$  & $R(17:2)$ \\
\hline
&&&&\\
$^{190}$Pb & 0.774 & 0.897 &$^{191}$Pb & 0.818   & 1.056       \\
$^{192}$Pb & 0.853 & 0.874 &$^{193}$Pb & 0.881   & 1.032  \\
$^{194}$Pb & 0.965 & 0.932 &$^{195}$Pb & 0.969   & 1.004      \\
$^{196}$Pb & 1.049 & 0.959 &$^{197}$Pb & 1.006   & 0.959  \\
\hline
\end{tabular}
\end{ruledtabular}
\label{tab:table2}
\end{center}
\end{table*}
We have further listed the experimental $E2$ gamma energies associated with the transitions $\Delta E^{2^+}_{0^+}$  and $\Delta E^{{15/2}^-}_{{11/2}^-}$ for the even-even and odd-A Sn-isotopes in Table 1~\cite{Jain2015, astier132, astier130,ENSDF,XUNDL}. The ratio of these transitions denoted as $R(15:2) = \Delta E^{{15/2}^-}_{{11/2}^-}$ / $\Delta E^{2^+}_{0^+}$ is observed to be $\sim$ 1 for the $^{114-125}$Sn isotopes. This suggests a complete alignment of the odd-neutron in the $h_{11/2}$ orbital, producing the ${11/2}^-$ spin state. This supports the observation that the ${11/2}^-$ state in odd-A Sn-isotopes and the $0^+$ state in the neighboring even-even Sn-isotopes have great similarity in their wave functions. Similarly, the observed $E2$ gamma transitions $\Delta E^{{12}^+}_{{10}^+}$ and $\Delta E^{{31/2}^-}_{{27/2}^-}$ in even-even and odd-A Sn-isotopes have also been listed in Table 1 for $^{118-125}$Sn-isotopes~\cite{astier132, astier130}. Fotiades et al.~\cite{fotiades11} have compared the almost identical energies and similar structure involved in the $\Delta E^{2^+}_{0^+}$  and $\Delta E^{{12}^+}_{{10}^+}$ $\gamma$-transitions within the same isotope for $^{116-126}$Sn, and suggested that the ${10}^+$ isomeric state comes from the two aligned neutrons in the $h_{11/2}$ orbital. We have calculated the ratio of the transitions in odd-A Sn-isotope and its even-even core Sn-isotope, denoted as $ R(31:12)= \Delta E^{{31/2}^-}_{{27/2}^-}$ / $\Delta E^{{12}^+}_{{10}^+}$, which is also observed to be $\sim$ 1. The known gamma transition energies for the $N=82$ isotones, have also been listed in Table 2~\cite{astier130}. The ratios $R(15:2)$ and $R(31:12)$ again have the value $\sim$ 1, wherever these could be obtained.
\begin{center}
\includegraphics[width=8cm,height=6cm]{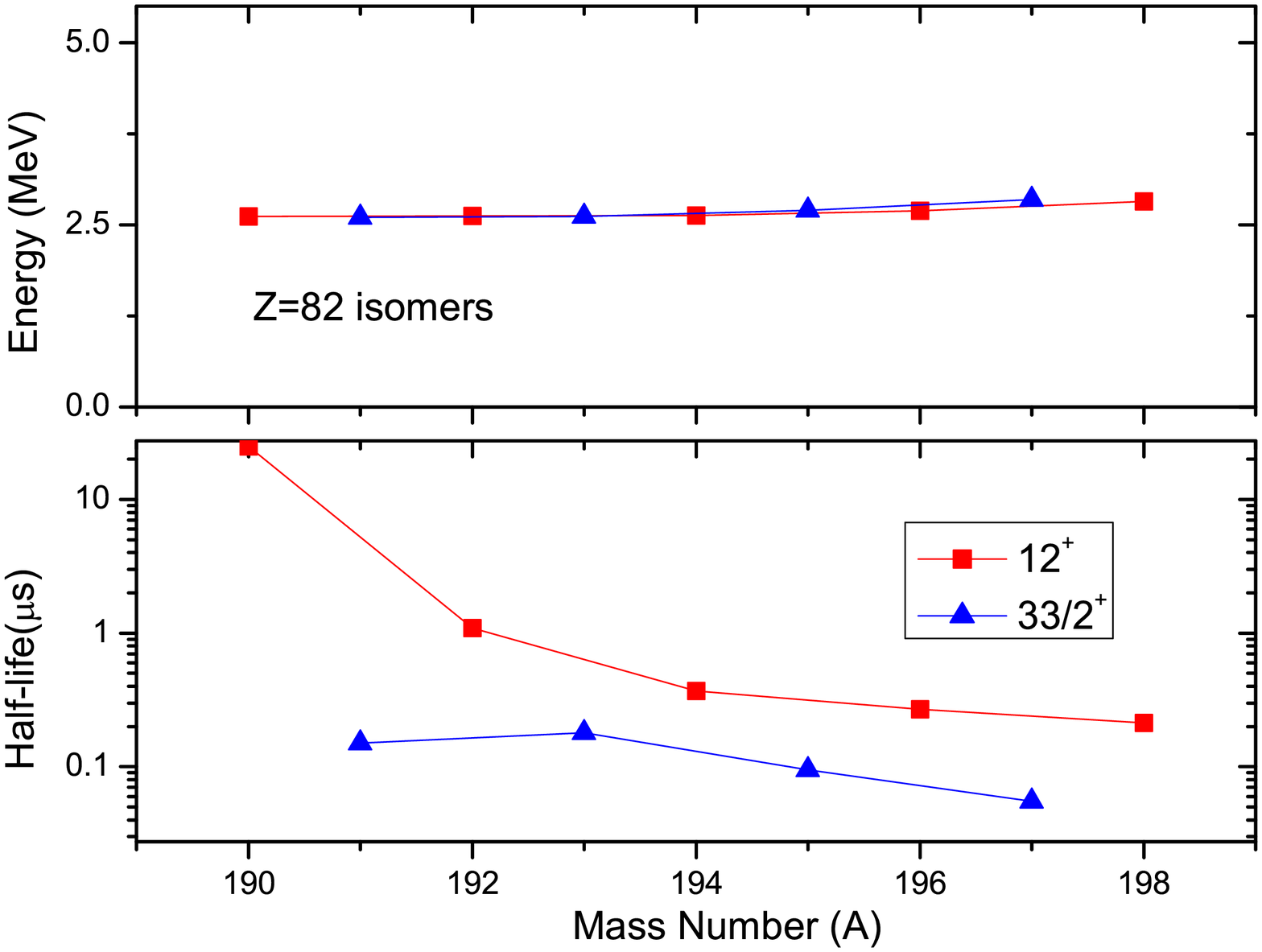}\\
Fig.\ 3\begin{minipage}[t]{69mm} \quad (color online)\,Variation of the experimental energy and half-life values of the ${12}^+$ and ${33/2}^+$ isomers in $Z=82$ isotopes. \\
\end{minipage}
\label{fig:halflifez50n82}
\end{center} 

It is obvious from the observed values that the seniority $\it{v}$ remains as $1$ for the ${11/2}^-$ states, as coming from the unique parity $h_{11/2}$ orbital of the active valence space. We may also infer from these observations that the ${10}^+$ and ${27/2}^-$ isomeric states, in both the chains, are maximally aligned decoupled states having similar wave functions and nuclear configurations. That is why the ${10}^+$ and ${27/2}^-$ states closely follow each other in excitation energy without exhibiting any odd-even effect, as shown in Fig. 1. We, therefore, foresee the seniority as $\it{v}$=0 for the $0^+$ states, $\it{v}$=1 for the ${11/2}^-$ states, $\it{v}$=2 for the $2^+$ states and $\it{v}$=3 for the ${15/2}^-$ states. Similarly, we may assign the seniority $\it{v}$=2 for the ${10}^+$ states, $\it{v}$=3 for the ${27/2}^-$ states, $\it{v}$=4 for the ${12}^+$ states, and $\it{v}$=5 for the ${31/2}^-$ states for these n-rich Sn isotopes. The same seniority difference $\Delta\it{v}$=2 between the ${15/2}^-$ and ${11/2}^-$ states, and for the $2^+$ and $0^+$ states gives their corresponding ratio $R(15:2)$ as $\sim$ 1. The difference $\Delta\it{v}$=2 also holds for the ${31/2}^-$ and ${27/2}^-$ states, and for the ${12}^+$ and ${10}^+$ states, which makes the ratio $R(31:12)$ $\sim$ 1.

We have also listed the experimental $E2$ gamma energies associated with the transitions $\Delta E^{2^+}_{0^+}$  and $\Delta E^{{17/2}^+}_{{13/2}^+}$ for the even-even and odd-A Pb-isotopes in Table 3~\cite{Jain2015,ENSDF,XUNDL}. The ratio of these transitions denoted as $R(17:2) = \Delta E^{{17/2}^+}_{{13/2}^+}$ / $\Delta E^{2^+}_{0^+}$ is observed to be $\sim$ 1 for the $^{190-197}$Pb isotopes. This suggests a complete alignment of the odd-neutron in the $i_{13/2}$ orbital, producing the ${13/2}^+$ spin state. This supports the observation that the ${13/2}^+$ state in odd-A Pb-isotopes and the $0^+$ state in the neighboring even-even Pb-isotopes have great similarity in their wave functions, very similar to the isomers in other two semi-magic chains. 

Similarly, the observed $E2$ gamma transitions $\Delta E^{{14}^+}_{{12}^+}$ have also been listed in Table 3 for the even-even $^{190-196}$Pb isotopes; however, no measurements are available for $\Delta E^{{37/2}^+}_{{33/2}^+}$ in odd-A $^{190-197}$Pb isotopes. We, therefore, could not calculated the ratio of the transitions in odd-A Sn-isotope and its even-even core Sn-isotope, denoted as $ R(37:14)= \Delta E^{{37/2}^+}_{{33/2}^+}$ / $\Delta E^{{14}^+}_{{12}^+}$, which is also expected to be $\sim$ 1. Therefore, similar arguments work for all the high-spin isomers in these $Z=50, N=82$ and $Z=82$ semi-magic chains, while their respective valence spaces and intruder orbitals are different. This similarity hints towards the goodness of seniority. Keeping our previous results for even-A Sn isotopes in mind~\cite{Maheshwari2016}, we have further done the generalized seniority calculations for decay rates to confirm the identical situation of various isomers from different nuclear regions. 

\section{Theoretical interpretation}

We briefly present the formulas used in the $B(E2)$ calculations using the generalized seniority scheme, and successfully applied to the ${10}^+$, and ${15}^-$ isomers in Sn isotopes in ~\cite{Maheshwari2016}. In this paper, we extend our studies to the ${10}^+$ isomers in $N=82$ isotones and the ${12}^+$ isomers in $Z=82$ isotopes. We further apply these results for the odd-A semi-magic nuclei, particularly for ${27/2}^-$, ${19/2}^+$, ${23/2}^+$ and ${35/2}^+$ isomers in Sn isotopes, ${27/2}^-$ isomers in $N=82$ isotones and ${33/2}^+$ isomers in $Z=82$ isotopes.

\subsection{$B(E2)$ rates from generalized seniority}

The $B(E2)$ values, between $J_i$ and $J_f$ states, in a mixed configuration $\tilde{j} = j \otimes j'....$  along with the corresponding total pair degeneracy $\Omega= \frac{1}{2}(2 \tilde{j} +1)= \frac{1}{2} \sum \limits_j (2j+1)$ by using the generalized seniority scheme can be written as follows
\begin{eqnarray}
B(E2)=\frac{1}{2J_i+1}|\langle \tilde{j}^n v l J_f || \sum_i r_i^2 Y^{2}(\theta_i,\phi_i) || \nonumber\\
\tilde{j}^n v' l' J_i \rangle |^2 
\end{eqnarray}
This implies that the $B(E2)$ values show a parabolic behavior in the multi$-j$ case depending upon the seniority of the states involved in the transition. We rewrite the seniority reduction formula for the reduced matrix elements with seniority conserving $\Delta v=0$ transitions between the initial and final states for the completeness of the text in paper. The relations are as follows:

\begin{eqnarray}
\langle \tilde{j}^n v l J_f ||\sum_i r_i^2 Y^{2}(\theta_i,\phi_i)|| \tilde{j}^n v l' J_i \rangle = \Bigg[ \frac{\Omega-n}{\Omega-v} \Bigg] \nonumber\\
\langle \tilde{j}^v v l J_f ||\sum_i r_i^2 Y^{2}(\theta_i,\phi_i)|| \tilde{j}^v v l' J_i \rangle
\end{eqnarray}

The $B(E2)$ values, which depends on the particle number $n$, the generalized seniority $v$ and the corresponding total pair degeneracy $\Omega$, can be calculated by using these formulas. These formulas take care of mixing of the active orbitals in the valence space. We present details of the calculations and results in the next section. Note that some information on radial integrals along with the matrix elements of spherical harmonics is hidden in the constant of proportionality. 

\subsection{Even-A Semi-magic Nuclei}

We have successfully shown that the high-spin $E2$ decaying ${10}^+$, ${15}^-$ and $E1$ decaying ${13}^-$ isomers are similar in their decay trends for even-even Sn isotopes, see paper ~\cite{Maheshwari2016} for details. We now apply the same formalism~\cite{Maheshwari2016} to the isomers in other semi-magic chains, particularly, the ${10}^+$ isomers in $N=82$ isotones, and the ${12}^+$ isomers in $Z=82$ isotopes. Both the isomeric chains decay by $E2$ transitions as shown in Fig. 5 and 6, and their measured $B(E2)$ trends are quite similar to the ${10}^+$ isomers in the Sn-isotopes (follow fig. 4). We choose the active valence space as $h_{11/2}$, $d_{3/2}$ and $s_{1/2}$ in the $N=82$ isotones and i$_{13/2}$, f$_{7/2}$ and p$_{3/2}$ in the Pb-isotopes. The resultant $\tilde{j}$ and $\Omega$ values become $17/2$ and $9$ in the $N=82$ isotones, similar as in the case of $Z=50$ isotopes, since both the $Z=50$ and $N=82$ chains share same orbitals in their active valence space of particle number $50-82$. We consider $Z=64$ as core by assuming the $g_{7/2}$ and $d_{5/2}$ orbitals as completely filled. 

Therefore, we fit the $n=2$ situation at $Z=66$, $^{146}$Dy for the seniority $v=2$ isomers in the $N=82$ chain, and get the resultant parabolic trend as shown in Fig. 5. The calculated trends reproduce the experimental data quite well; $\Omega=6$ gives the best fit to the data, a different situation as compared to the ${10}^+$, Z=50 isomers, where $\Omega=9$ gives the best fit. This can be understood in terms of different valence particles, particularly the involvement of protons in the case of N=82 isotones. It appears to be closer to the pure seniority scheme, as quoted in the previous literature ~\cite{talmi93}. 

\begin{center}
\includegraphics[width=8cm,height=6cm]{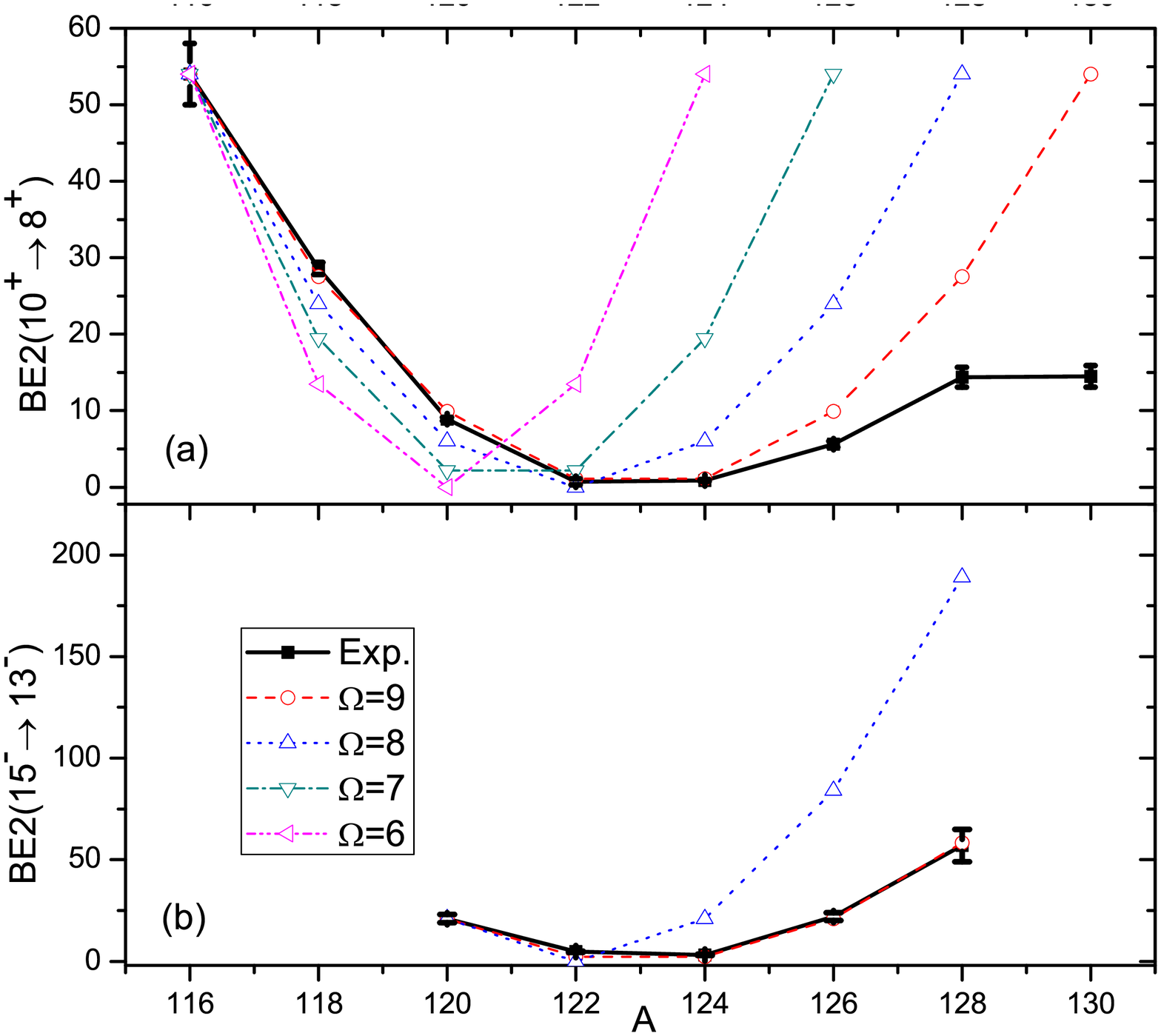}\\
Fig.\ 4\begin{minipage}[t]{69mm} \quad (color online)\,Variation of the $B(E2)$ values of the ${10}^+$ and ${15}^-$ isomers in $Z=50$ isotopes~\cite{Maheshwari2016}. All the values are shown in the units of $e^2fm^4$. \\
\end{minipage}
\label{fig:z50}
\end{center}

On the other hand, the resultant $\tilde{j}$ and $\Omega$ values become $25/2$ and $13$ for the Pb-isotopes, where $\tilde{j} = i_{13/2} \otimes f_{7/2} \otimes p_{3/2}$. The active valence space of $N=82-126$ for these isotopes consists of $h_{9/2}$, $i_{13/2}$, $f_{7/2}$, $p_{3/2}$, $f_{5/2}$ and $p_{1/2}$ orbitals. We consider $h_{9/2}$ as completely filled, so the next three active orbitals $i_{13/2}$, $f_{7/2}$ and $p_{3/2}$ will be filled at $^{200}$Pb. We fix the proportionality constants by fitting the measured values for $^{198}$Pb for the seniority $v=2$ isomers in the $Z=82$ chains, respectively. The calculated results reproduce the experimental trends for both these isomeric chains as shown in Fig. 6. The calculations have been done by considering the transitions as seniority conserving ones ($\Delta v=0$) along with generalized seniority of these states as $v=2$. One can also expect the occurrence of these generalized seniority $v=2$ isomers in the gaps of experimental data for both the $N=82$ and $Z=82$ chains extending towards proton-deficient and neutron-deficient sides, respectively (follow the figs. 5 and 6 for the same).

\begin{center}
\includegraphics[width=8cm,height=5cm]{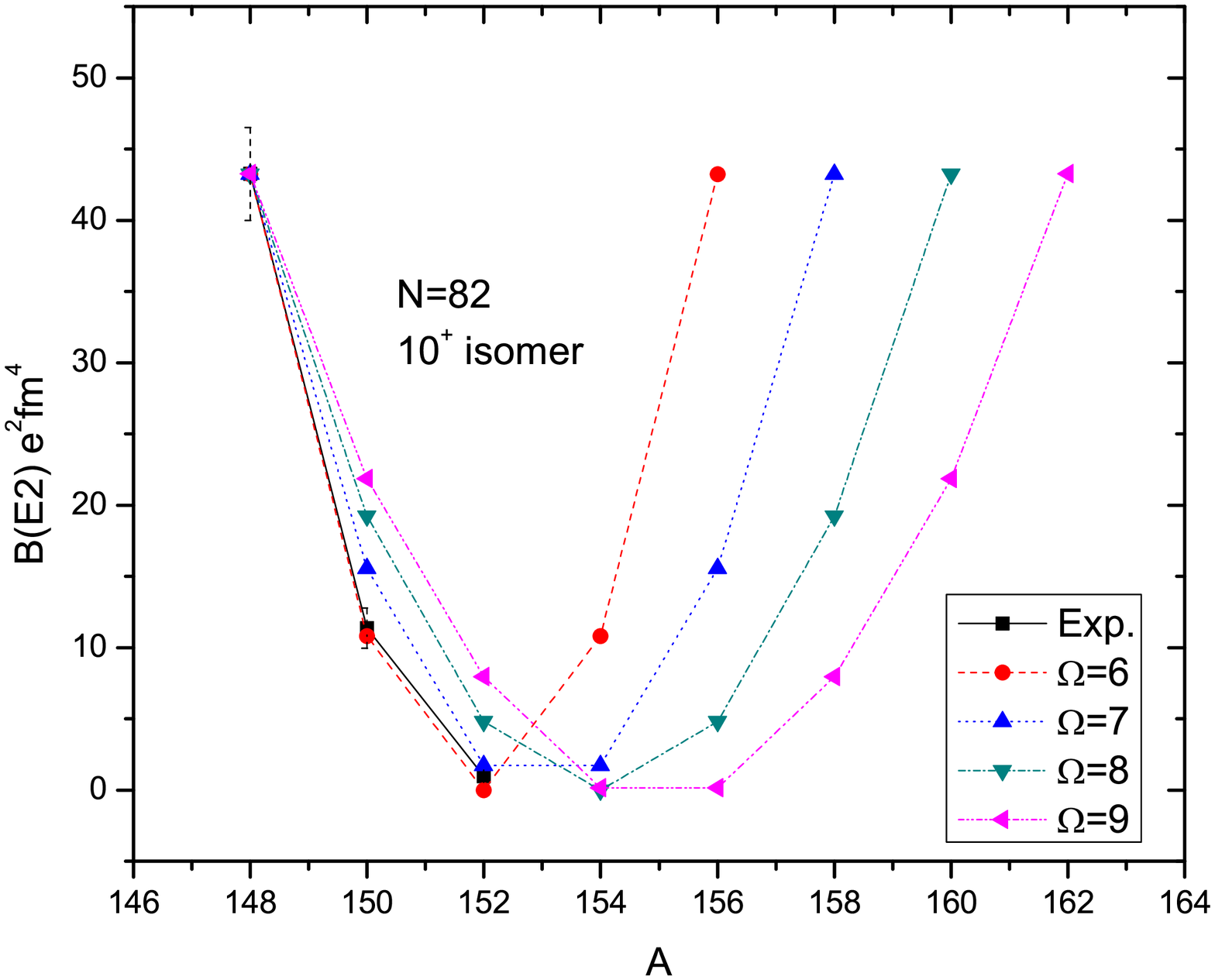}\\
Fig.\ 5\begin{minipage}[t]{69mm} \quad (color online)\,Variation of the $B(E2)$ values of the ${10}^+$ isomers in $N=82$ isotones.\\
\end{minipage}
\label{fig:n82}
\end{center}

\begin{center}
\includegraphics[width=8cm,height=5cm]{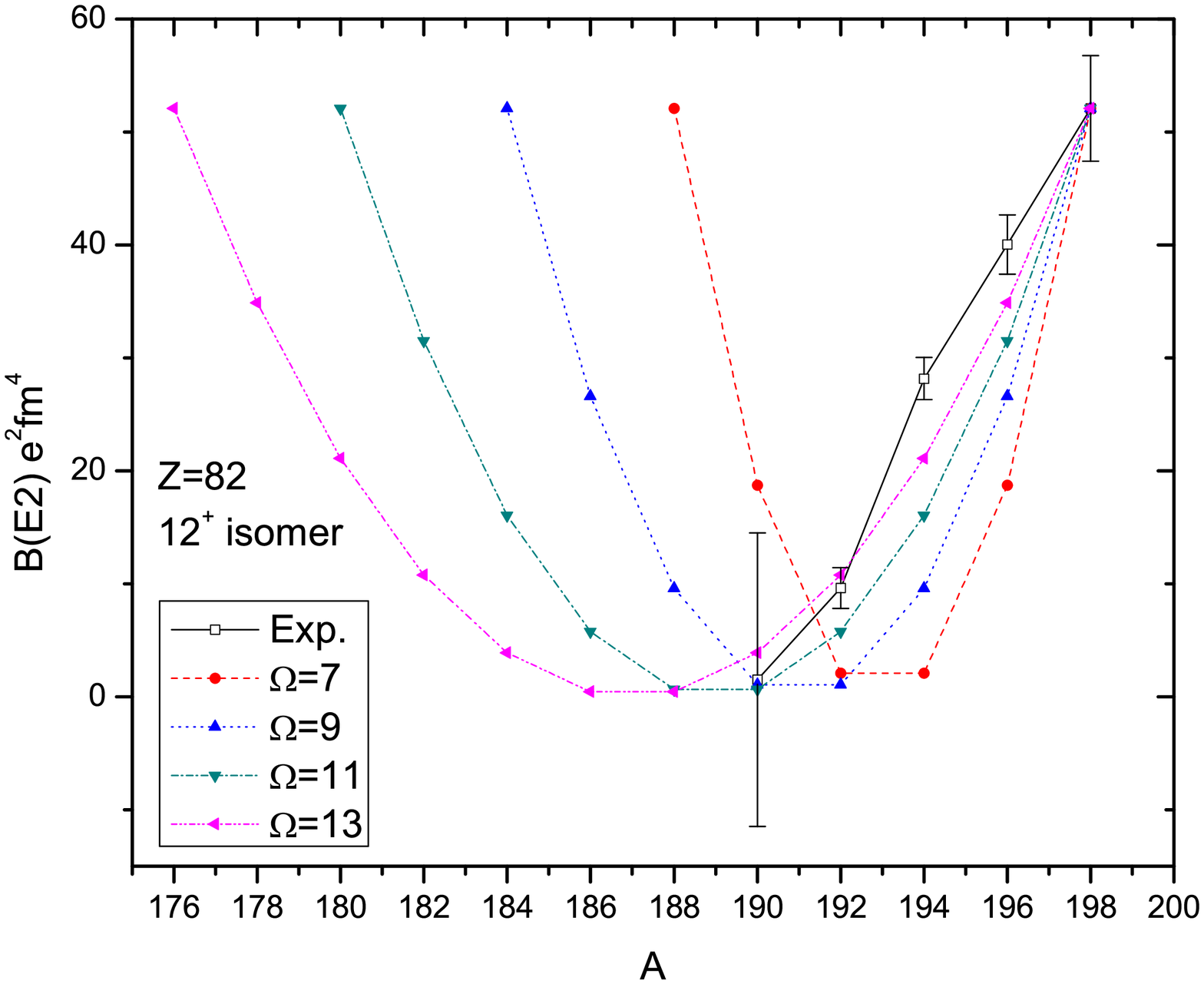}\\
Fig.\ 6\begin{minipage}[t]{69mm} \quad (color online)\,Variation of the $B(E2)$ values of the ${12}^+$ isomers in Pb-isotopes.\\
\end{minipage}
\label{fig:z82}
\end{center}

One can, therefore, observe the identical behavior of the $B(E2)$s in the high-spin isomers for all the three semi-magic chains, $Z=50$, $N=82$, and $Z=82$. This is due to their identical generalized seniorities and the $\Delta v=0$ transitions. It is interesting to note that the valence particles in $Z=50$ chain are neutrons while protons become active in the case of $N=82$ chain, and they share same orbitals in the active valence space of $50-82$. On the other hand, the $Z=82$ chain has a different neutron valence space along with different set of active orbitals. In spite of these differences with each other, high-spin isomers in all the three chains appear to follow the same microscopic scheme of generalized seniority. This highlights the importance of configuration mixing required in the generation of these states. Hence, the generalized seniority behaves almost as a good quantum number for these states in all three semi-magic chains. 

We note that the seniority $v=4$ isomers are only known in the $Z=50$ isotopes up to now. Due to the strong validity of generalized seniority in all the three semi-magic chains, one can expect and predict the high-spin and high-seniority $v=4$ isomers in the remaining two chains as well. Measurements in this direction should be made to confirm this scenario.

\subsection{Odd-A Semi-magic Nuclei}

We now study the ${27/2}^-$, ${19/2}^+$ and ${23/2}^+$ isomers of odd-A Sn-isotopes in light of the generalized seniority scheme. We assume that the $g_{7/2}$ and $d_{5/2}$ orbitals are completely filled up to $^{114}$Sn; hence, the remaining active orbitals for the Sn-isomers are $h_{11/2}$, $d_{3/2}$ and $s_{1/2}$ orbitals in the 50$−$82 valence space. We have performed the generalized seniority calculations assuming $v=3$, and $\Delta v=0$ transitions and by fitting the value of $^{119}$Sn isotope, as shown in Fig. 7. We calculate the $B(E2)$ values for the ${27/2}^-$ isomers using $\Omega$ values of $7$, $8$ and $9$ corresponding to $\tilde{j} = h_{11/2} \otimes s^2_{1/2}$, $\Omega = 7$; $\tilde{j} = h_{11/2} \otimes d^2_{3/2}$, $\Omega = 8$; $\tilde{j} = h_{11/2} \otimes d_{3/2} \otimes s_{1/2}$, $\Omega = 9$, respectively. We present the calculated and experimental $B(E2)$ values for the ${27/2}^-$ isomers in Fig. 7 for comparison. One can see that the calculated values for $\Omega=9$ fit the experimental data reasonably well. This confirms that the ${27/2}^-$ isomers behave as generalized seniority $v=3$ isomers having $\Delta v=0$ transitions to the lower lying ${23/2}^-$ states, and support the mixing of all the three $h_{11/2}$, $d_{3/2}$ and $s_{1/2}$ orbitals. 

We also plot the $B(E2)$ values vs. A (mass number) for the ${19/2}^+$ and ${23/2}^+$ isomers, in the top and bottom panels of Fig. 8, respectively. Experimental data for these isomers have been taken from the recent measurements of Iskra et al.~\cite{Iskra2016} and the references therein. We find that the calculated results from $\Omega=9$ value, and generalized seniority $v=3$ with $\Delta v=0$ transitions, are again able to explain the experimental trend reasonably well. Note that we fit the experimental value of $^{129}$Sn in calculations for both the ${19/2}^+$ and ${23/2}^+$ isomers. This confirms that the ${19/2}^+$, ${23/2}^+$ isomers decay to the same seniority states, and are generalized seniority $v=3$ isomers (involving $h_{11/2}$, $d_{3/2}$ and $s_{1/2}$ orbitals). Note that the seniority $v=4$, ${13}^-$, ${15}^-$ isomers, and seniority $v=2$, ${10}^+$ isomers in the even-even $^{116-130}$Sn isotopes have already been explained by using the same mixed configuration having $\Omega$ value of $9$~\cite{Maheshwari2016}. We can, therefore, conclude that the high-spin isomers in $^{116-130}$Sn mass region arise from the mixing of all the available $h_{11/2}$, $d_{3/2}$ and $s_{1/2}$ orbitals for both even and odd-A isotopes. 

On the other hand, we find that the inclusion of $d_{5/2}$ orbital along with $h_{11/2}$, $d_{3/2}$ and $s_{1/2}$ orbitals is required for explaining the first $2^+$ states in these Sn-isotopes~\cite{Maheshwari20161}. However, the configuration changes for the first excited $3^-$ states showing the octupole character having $d-h$ orbitals~\cite{unpublished}. It is quite obvious that the dominance of $h_{11/2}$ orbital increases, while going towards the high-spin and high-seniority states. This type of information by generalized seniority guides us to infer the nature of configuration mixing and wave functions involved in the generation of a given set of states. Also, the structure information for both even- and odd-A nuclei are similar, irrespective of the mass region, involved valence particles, orbitals, etc., except for an extra nucleon in odd-A systems. To sum up, we find that the seniority and generalized seniority provides a finger-print evidence for the similarity between various isomers in the semi-magic chains.

\begin{center}
\includegraphics[width=8cm,height=5cm]{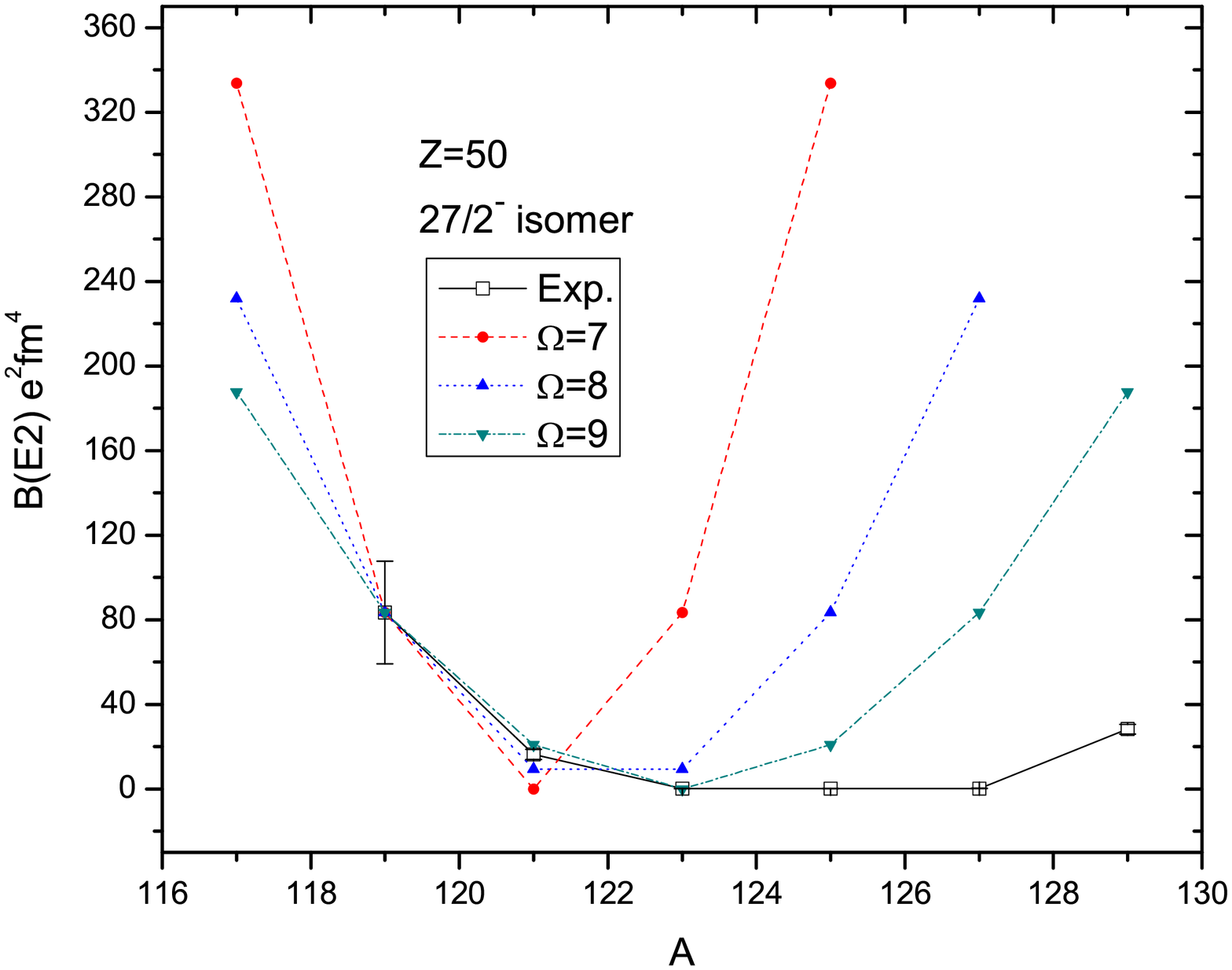}\\
Fig.\ 7\begin{minipage}[t]{69mm} \quad (color online)\,Variation of the $B(E2)$ values of the ${27/2}^-$ isomers in Sn-isotopes~\cite{unpublished}. \\
\end{minipage}
\label{fig:isomer27}
\end{center}

On the basis of this interpretation, we have also analyzed the single measured value at $^{123}$Sn, for the higher seniority $v=5$, $({35/2}^+)$ isomer. We have fitted the value of $^{123}$Sn, and calculated the values for the other neighboring isotopes assuming $\Delta v=0$ transitions using $\Omega=8$ and the corresponding mixed configuration. We have taken $^{114}$Sn as core; this implies that the first location $(n=1)$ to have seniority $v=5$ state corresponds to $^{119}$Sn. The $\Omega=9$ and the related mixed configuration can not be fitted using the experimental value at $^{123}$Sn as it leads to a zero value for the coefficient ${(\frac{\Omega-n}{\Omega-v}})^2$ in the middle($n=v=5$). We have plotted in Fig. 9, the $B(E2)$ trend for the seniority $v=5$, ${35/2}^+$ isomer using $\Omega=8$ value and the respective mixed configuration. These calculations, therefore, help us to predict some unknown values also. It is quite obvious that these high-seniority $v=5$, ${35/2}^+$ isomers in odd-A n-rich Sn-isotopes can be related to the seniority $v=4$ isomers in even-A n-rich Sn-isotopes. 

\begin{center}
\includegraphics[width=8cm,height=6cm]{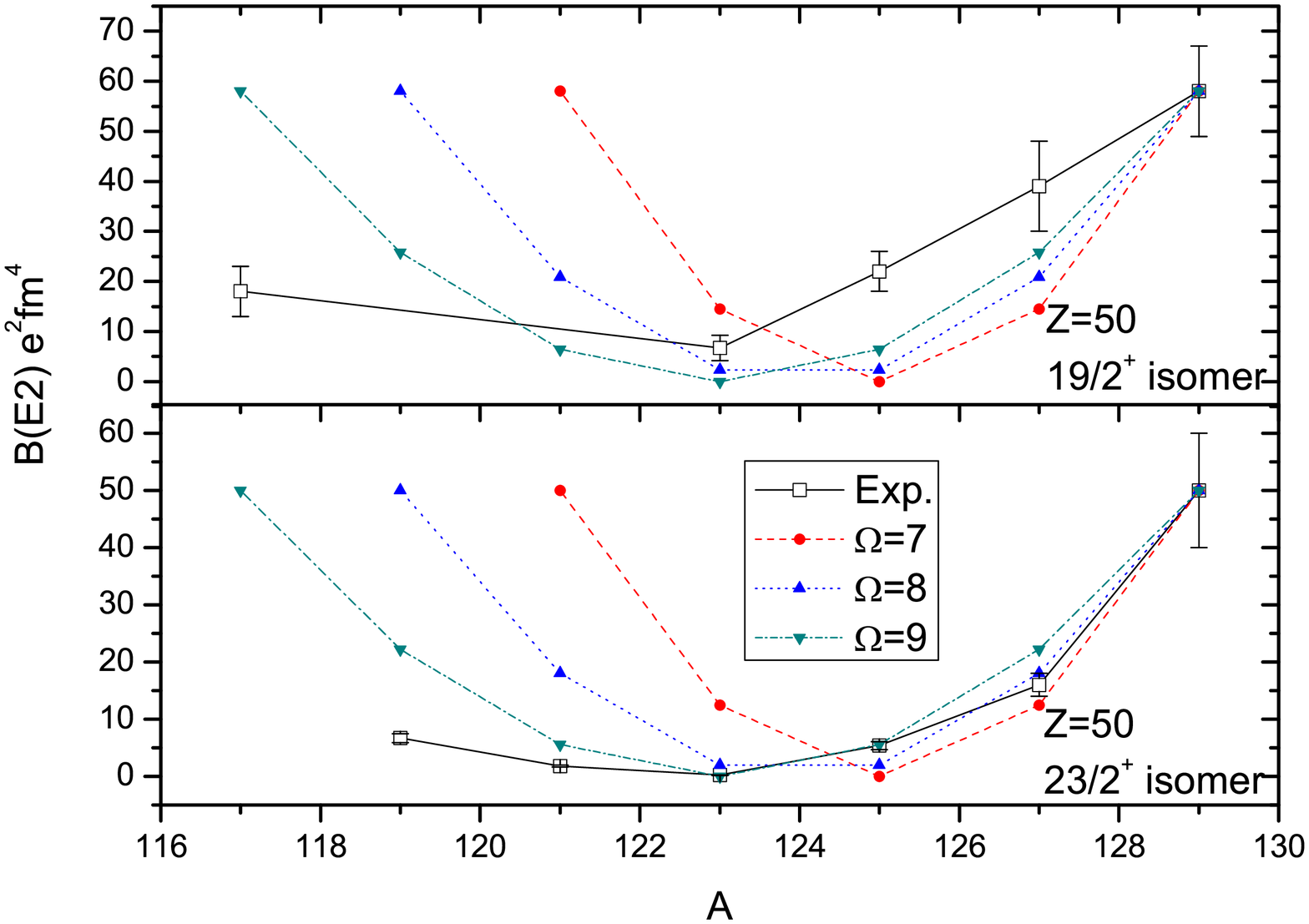}\\
Fig.\ 8\begin{minipage}[t]{69mm} \quad (color online)\,Variation of the $B(E2)$ values of the ${19/2}^+$, and ${23/2}^+$ isomers in Sn-isotopes~\cite{unpublished}. \\
\end{minipage}
\label{fig:isomer1923}
\end{center}

We further study the ${27/2}^-$ isomers in odd-A $N=82$ isotones using the same generalized seniority scheme. We have already pointed out that the Sn isotopes and $N=82$ isotones have same set of orbitals in the active valence space of nucleon number $50-82$, while neutrons are active in Sn isotopes and protons are active in the $N=82$ isotonic chain. We assume that the $g_{7/2}$ and $d_{5/2}$ orbitals are completely filled up to $^{146}$Gd (Z=64); hence, the remaining active orbitals for these $N=82$ isomers are $h_{11/2}$, $d_{3/2}$ and $s_{1/2}$ orbitals. We have performed the generalized seniority calculations assuming $v=3$, and $\Delta v=0$ transitions and by fitting the value of $^{149}$Ho isotone, as shown in Fig. 10. We calculate the $B(E2)$ values for the ${27/2}^-$ isomers using $\Omega$ values of $7$, $8$ and $9$ corresponding to $\tilde{j} = h_{11/2} \otimes s^2_{1/2}$, $\Omega = 7$; $\tilde{j} = h_{11/2} \otimes d^2_{3/2}$, $\Omega = 8$; $\tilde{j} = h_{11/2} \otimes d_{3/2} \otimes s_{1/2}$, $\Omega = 9$, respectively, as in the case of Sn isotopes. We present the calculated and experimental $B(E2)$ values for the ${27/2}^-$ isomers in Fig. 10 for comparison. One can see that the calculated values for $\Omega=6$ best fit the experimental data. This confirms that the ${27/2}^-$ isomers behave as more like pure seniority $v=3$ isomers having $\Delta v=0$ transitions to the lower lying ${23/2}^-$ states, and support the dominance of $h_{11/2}$ orbital. The same argument has already been shown to work for even-A N=82 isotones too. 

We next present a comparison of the ${33/2}^+$ isomers in Pb-isotopes on the same footing, where we again calculate the $B(E2)$ values for these isomers using $i_{13/2}$, $f_{7/2}$ and $p_{3/2}$ as active orbitals. We use the possible mixed configurations: $\Omega=7$ ($i_{13/2}$),  $\Omega=9$ ($i_{13/2} \otimes p^2_{3/2}$), $\Omega=11$ ($i_{13/2} \otimes  f^2_{7/2}$), and $\Omega=13$ $(i_{13/2} \otimes  p_{3/2} \otimes f_{7/2})$. We assume that the lowest lying $h_{9/2}$ orbital is full, and therefore, the other three orbitals $(i_{13/2} \otimes  p_{3/2} \otimes f_{7/2})$ become full at $^{200}$Pb. Since we do not have experimental data at lower mass neutron-deficient side, we fit the value of $^{197}$Pb by assuming $^{200}$Pb as the completely full configuration, using generalized seniority $v=3$ for $\Delta v=0$ transitions. The calculated values from $\Omega =13$ having mixing of all the three active orbitals are able to explain the experimental data quite closely (See Fig. 11).

\begin{center}
\includegraphics[width=8cm,height=5cm]{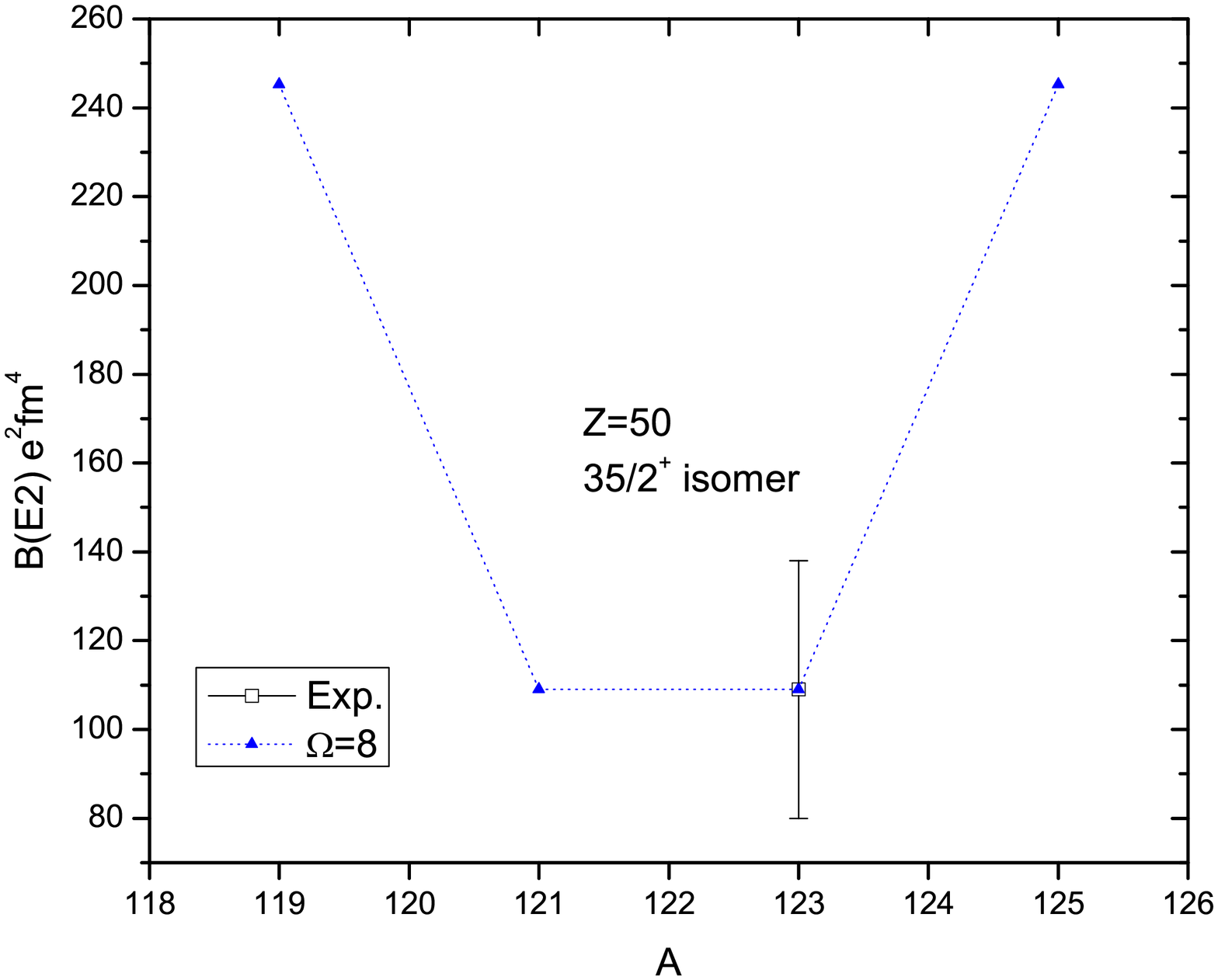}\\
Fig.\ 9\begin{minipage}[t]{69mm} \quad (color online)\,Variation of the $B(E2)$ values of the ${35/2}^+$ isomers in Sn-isotopes~\cite{unpublished}. \\
\end{minipage}
\label{fig:isomer35}
\end{center}

\begin{center}
\includegraphics[width=8cm,height=5cm]{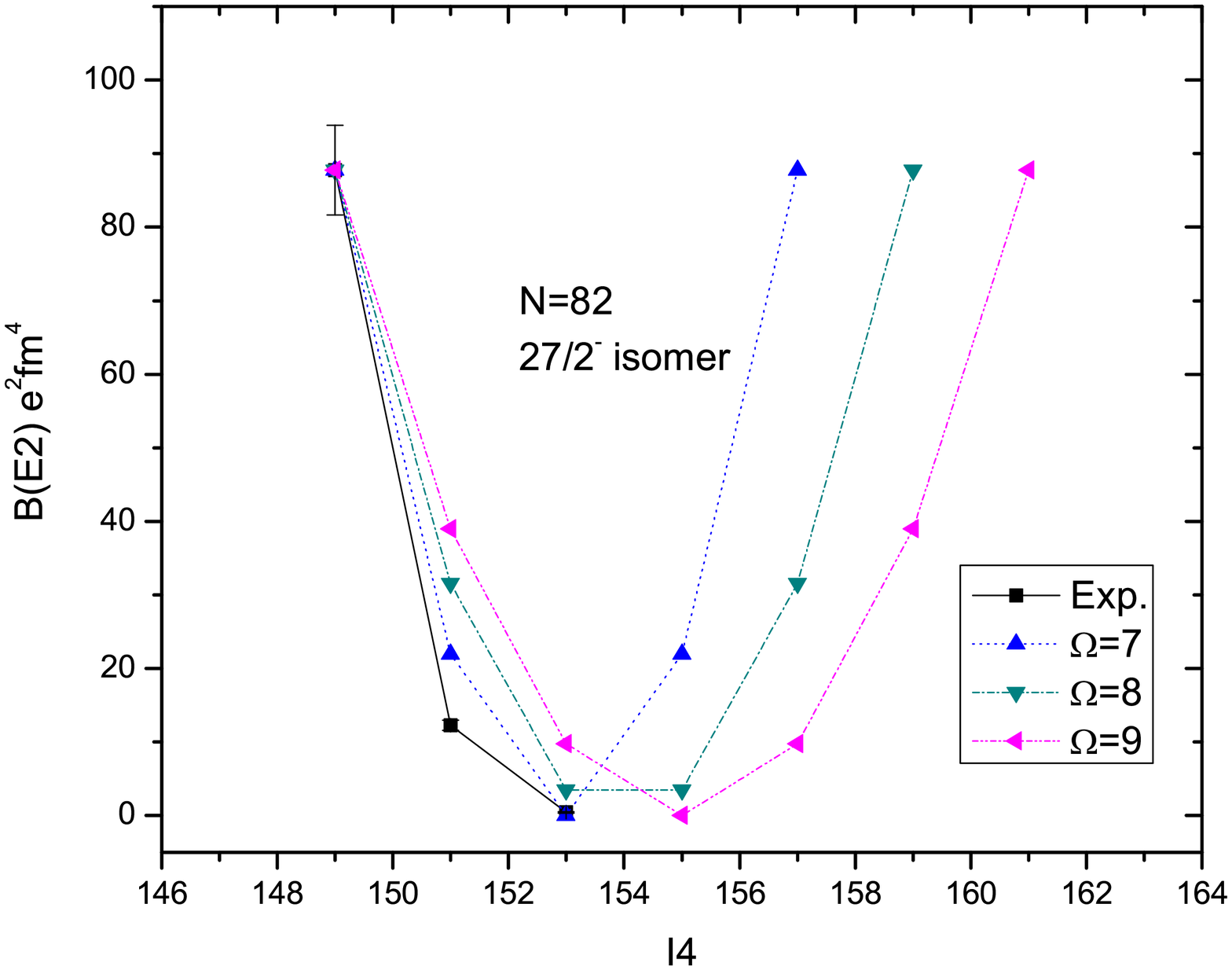}\\
Fig.\ 10\begin{minipage}[t]{69mm} \quad (color online)\,Variation of the $B(E2)$ values of the ${10}^+$ isomers in $N=82$ isotones.\\
\end{minipage}
\label{fig:n821}
\end{center}

We, therefore, conclude that the ${33/2}^+$ isomers are seniority $v=3$ isomers in Pb-isotopes, similar to the ${27/2}^-$ isomers in Sn-isotopes. Their similar $B(E2)$ trends can be explained in terms of the involvement of same seniority and generalized seniority, though coming from the different orbitals and different valence spaces. However, more measurements are required to obtain experimental data for the remaining nuclei and complete the picture. The same argument has been presented for the isomers in even-A nuclei. The seniority and generalized seniority hence play a unifying role in explaining the similar behavior in different sets of semi-magic nuclei. 

\begin{center}
\includegraphics[width=8cm,height=5cm]{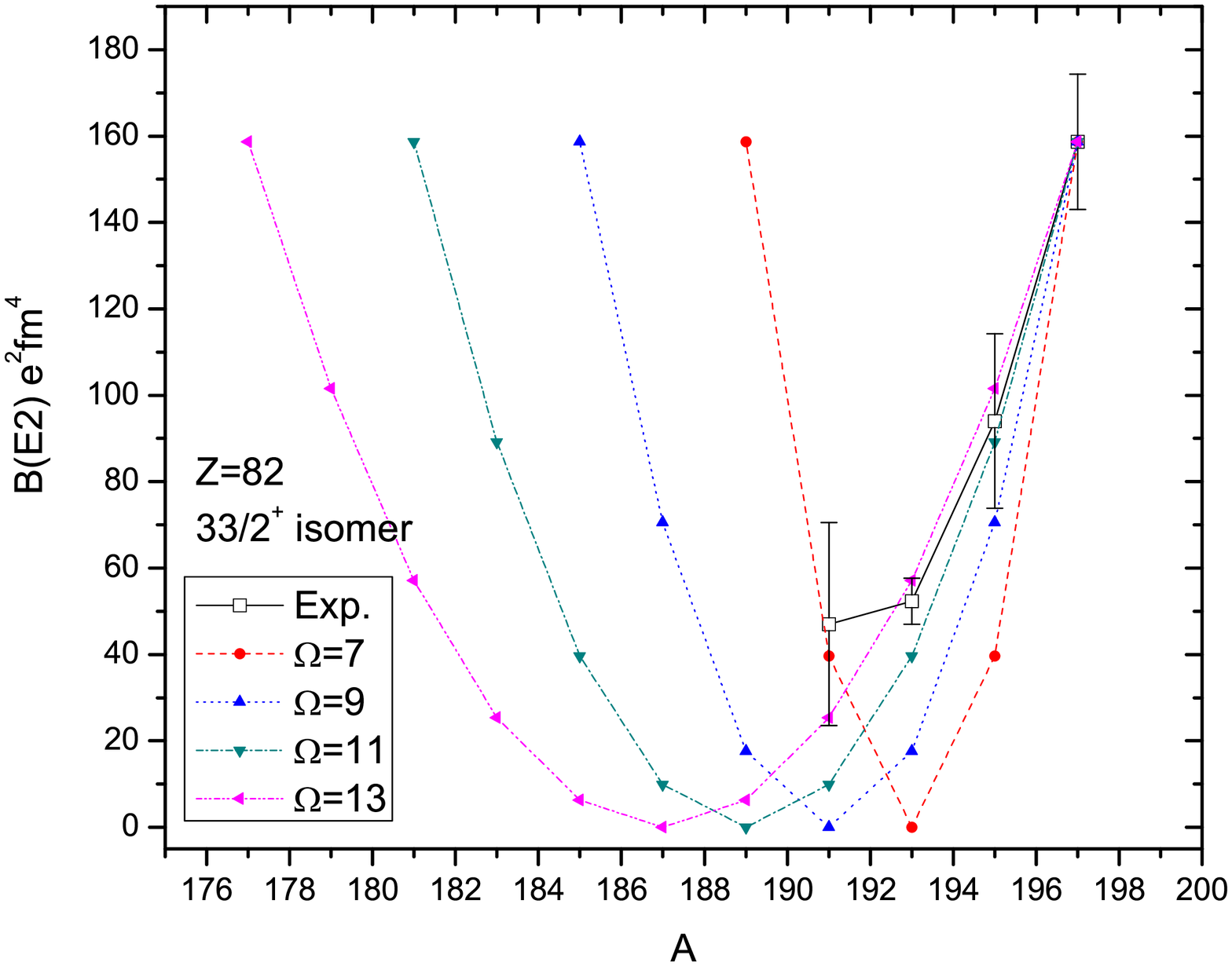}\\
Fig.\ 11\begin{minipage}[t]{69mm} \quad (color online)\,Variation of the $B(E2)$ values of the ${33/2}^+$ isomers in Pb-isotopes~\cite{unpublished}. \\
\end{minipage}
\label{fig:isomer33}
\end{center}

\section{Conclusion}

We have used the quasi-spin formalism for degenerate multi$-j$ orbitals to calculate the reduced electric transition probabilities in the semi-magic isomers. We find that the configuration mixing is essential to fully describe the $v=2$, ${10}^+$ isomers in the even-A Sn-isotopes, and the $v=3$, ${27/2}^-$, ${19/2}^+$, ${23/2}^+$ and $v=5$, ${35/2}^+$ isomers in the odd-A Sn-isotopes and to explain the $B(E2)$ values in all these Sn-isomers. This formalism reproduces the experimental trend quite well and is also capable to predict some numbers for the gaps in the measurements. On the other hand, the situation for the $v=2$, ${10}^+$ and the $v=3$, ${27/2}^-$ isomers in $N=82$ chain becomes different and highlights the dominance of $h_{11/2}$ orbital only (pure-seniority scheme). These isomers have also been compared with the identical trends for the $v=2$, ${12}^+$ isomers, and the $v=3$, ${33/2}^+$ isomers in the $Z=82$ chain. The identical behavior of all the high-spin isomers in various semi-magic chains strongly supports the goodness of seniority and generalized seniority up to very high-spin. This simple scheme of calculating the $B(EL)$ values may also be used to estimate the half-lives in unknown cases and, hence predict new isomers.

\renewcommand\refname{{\normalsize\bf References:}}

\end{document}